\documentclass[aps,prc,twocolumn,longbibliography,superscriptaddress,amsmath,nofootinbib]
{revtex4-1}

\usepackage[english]{babel}
\usepackage[utf8]{inputenc}
\usepackage[colorlinks, urlcolor=blue,linkcolor=blue, anchorcolor=blue, citecolor=blue]{hyperref}

\usepackage{multirow}
\usepackage{array}
\usepackage{amsthm}
\usepackage{mathtools}
\usepackage{physics}
\usepackage{xcolor}
\usepackage{graphicx}
\usepackage{bm}
\usepackage{soul}
\usepackage{adjustbox}
\usepackage{placeins}
\usepackage[T1]{fontenc}
\usepackage{lipsum}
\usepackage{csquotes}
\newcommand{\snn}{\sqrt{s_{NN}}}
\newcommand{\gev}{\textrm{GeV}}
\newcommand{\xperp}{\bm{x}_\perp}

\usepackage{soul}

\begin{document}


\title{Probing initial baryon stopping and equation of state with rapidity-dependent\\ directed flow of identified particles}

\author{Lipei Du}
\affiliation{Department of Physics, McGill University, Montreal, Quebec H3A 2T8, Canada}
\author{Chun Shen}
\affiliation{Department of Physics and Astronomy, Wayne State University, Detroit, Michigan 48201, USA}
\affiliation{RIKEN BNL Research Center, Brookhaven National Laboratory, Upton, New York 11973, USA}
\author{Sangyong Jeon}
\affiliation{Department of Physics, McGill University, Montreal, Quebec H3A 2T8, Canada}
\author{Charles Gale}
\affiliation{Department of Physics, McGill University, Montreal, Quebec H3A 2T8, Canada}
\date{\today} 

\begin{abstract}
Using a (3+1)-dimensional hybrid framework with parametric initial conditions, we study the rapidity-dependent directed flow  $v_1(y)$ of identified particles, including pions, kaons, protons, and lambdas in heavy-ion collisions. Cases involving Au+Au collisions are considered, performed at $\sqrt{s_{\rm NN}}$ ranging from 7.7 to 200 GeV. The dynamics in the beam direction is constrained using the measured pseudo-rapidity distribution of charged particles and the net proton rapidity distribution. Within this framework, the directed flow of mesons is driven by the sideward pressure gradient from the tilted source, and that of baryons mainly due to the initial asymmetric baryon distribution with respect to the beam axis driven by the transverse expansion. Our approach successfully reproduces the rapidity- and beam energy-dependence of $v_1$ for both mesons and baryons. We find that the $v_1(y)$ of baryons has strong constraining power on the initial baryon stopping, and together with that of mesons, the directed flow probes the equation of state of  the dense nuclear matter at finite chemical potentials. 
\end{abstract}

\maketitle


{\it \textbf{Introduction.}---}
Heavy-ion collision experiments performed at various beam energies constitute the main method for the exploration and mapping of the quantum chromodynamics (QCD) phase diagram \cite{Bzdak:2019pkr}. The dynamical evolution of collisions can span several different phases, and multistage hybrid frameworks have been developed for the theoretical modeling of that evolution. Model-to-data comparisons have confirmed the success of these hybrid approaches, and properties of the produced dense nuclear matter have been extracted quantitatively using Bayesian inference schemes. Among the observables, the collective flow of the final emitted particles can provide insight into the early-stage dynamics, transport properties, and equation of state (EoS)  of heavy-ion collisions \cite{Herrmann:1999wu,Stoecker:2004qu}. 

The directed flow, which describes the collective sideward motion, is believed to develop in the early stage of the collisions and thus reflects the EoS at those early times \cite{Sorge:1996pc,Herrmann:1999wu,Stoecker:2004qu}. The directed flow is defined as the first Fourier component in the azimuthal angle distribution of the final particles with respect to the reaction plane angle $\Psi$, $v_1=\langle\cos(\phi-\Psi)\rangle$.  Measurements of the rapidity-dependent $v_1(y)$ of identified particles have been carried out for several beam energies and colliding systems  at the BNL Alternating Gradient Synchrotron (AGS), the CERN Super Proton
Synchrotron (SPS), and the BNL Relativistic Heavy Ion Collider (RHIC) \cite{E895:2000maf,NA49:2003njx,PHOBOS:2005ylx,STAR:2005btp,STAR:2008jgm,STAR:2011hyh,STAR:2014clz,STAR:2017okv}. It is common practice to present the strength of $v_1(y)$ using its slope near midrapidity, $dv_1/dy|_{y=0}$. At AGS and SPS energies, $dv_1/dy|_{y=0}$ is observed to be positive for protons and negative for pions \cite{E895:2000maf,NA49:2003njx}. 

 At RHIC energies, and at intermediate impact parameters, $dv_1/dy|_{y=0}$ for protons changes sign from positive to negative between 7.7 and 11.5 GeV, with a minimum between 11.5 and 19.6 GeV, while for pion it is negative at all measured energies \cite{STAR:2005btp,STAR:2008jgm,STAR:2011hyh,STAR:2014clz,STAR:2017okv}. Within a large rapidity window covering the fragmentation regions, $dv_1/dy$ at midrapidity may differ from the regions closer to beam rapidity, and thus $v_1(y)$ may exhibit a cubic rapidity dependence where $dv_1/dy$ changes sign twice.
 
It has been argued that the non-monotonic collision energy dependence of $dv_1/dy|_{y=0}$ is a signature of the first-order phase transition of the quark-gluon plasma, especially if it is observed for baryons \cite{Brachmann:1999xt,Csernai:1999nf,Stoecker:2004qu}. A softening of the EoS \cite{Hung:1994eq,Rischke:1995pe} attributed to a first-order phase transition between hadronic matter and quark-gluon plasma at high baryon density region  \cite{Rischke:1995pe,Csernai:1999nf,Brachmann:1999xt,Stoecker:2004qu} can cause a significant reduction of the directed flow during the expansion. Thus a minimum in $dv_1/dy|_{y=0}$ as a function of beam energy may indicate such a first-order phase transition (the ``softest-point'' effect \cite{Stoecker:2004qu}). However, $v_1(y)$ is also strongly sensitive to other dynamical aspects of heavy-ion collisions unrelated to softening features of the EoS. For example, some qualitative features of the proton's $v_1(y)$ can be explained in a purely hadronic picture using strong space-momentum correlations together with strong but incomplete baryon stopping \cite{Snellings:1999bt}. The negative slope of pions can be explained by emission from a tilted source with respect to the beam direction \cite{Bozek:2010bi,Chatterjee:2017ahy,Bozek:2022svy}. Although various models have been used to calculate $v_1(y)$ from AGS to top RHIC energies, including transport models \cite{Bleicher:2000sx,Isse:2005nk,Guo:2012qi,Konchakovski:2014gda,Nara:2016hbg,Zhang:2018wlk,Nara:2020ztb,Nara:2021fuu}, three-fluid dynamics models \cite{Ivanov:2005yw,Konchakovski:2014gda,Ivanov:2014ioa}, and hydrodynamic models \cite{Steinheimer:2014pfa,Shen:2020jwv,Jiang:2021ajc,Ryu:2021lnx},
none of the existing dynamical models explain the main features of the beam energy dependence of the directed flow. Furthermore, different models yield results varying widely over the span of the available beam energies and collision systems \cite{Singha:2016mna,Nara:2021fuu}. 

In this work, we shall study the $v_1(y)$ for mesons and baryons, using a (3+1)-dimensional hybrid framework \cite{suppl} with parametric initial conditions. We explore how reproducing the measured flow constrains initial conditions, especially the baryon density and thus the initial baryon stopping, together with the measured pseudo-rapidity distribution of charged particles and the net proton rapidity distribution. We also study how the rapidity dependence of the directed flow of mesons and baryons probes the EoS of dense nuclear matter at finite chemical potential.

\begin{figure}[!tb]
\begin{center}
\includegraphics[width=0.94\linewidth]{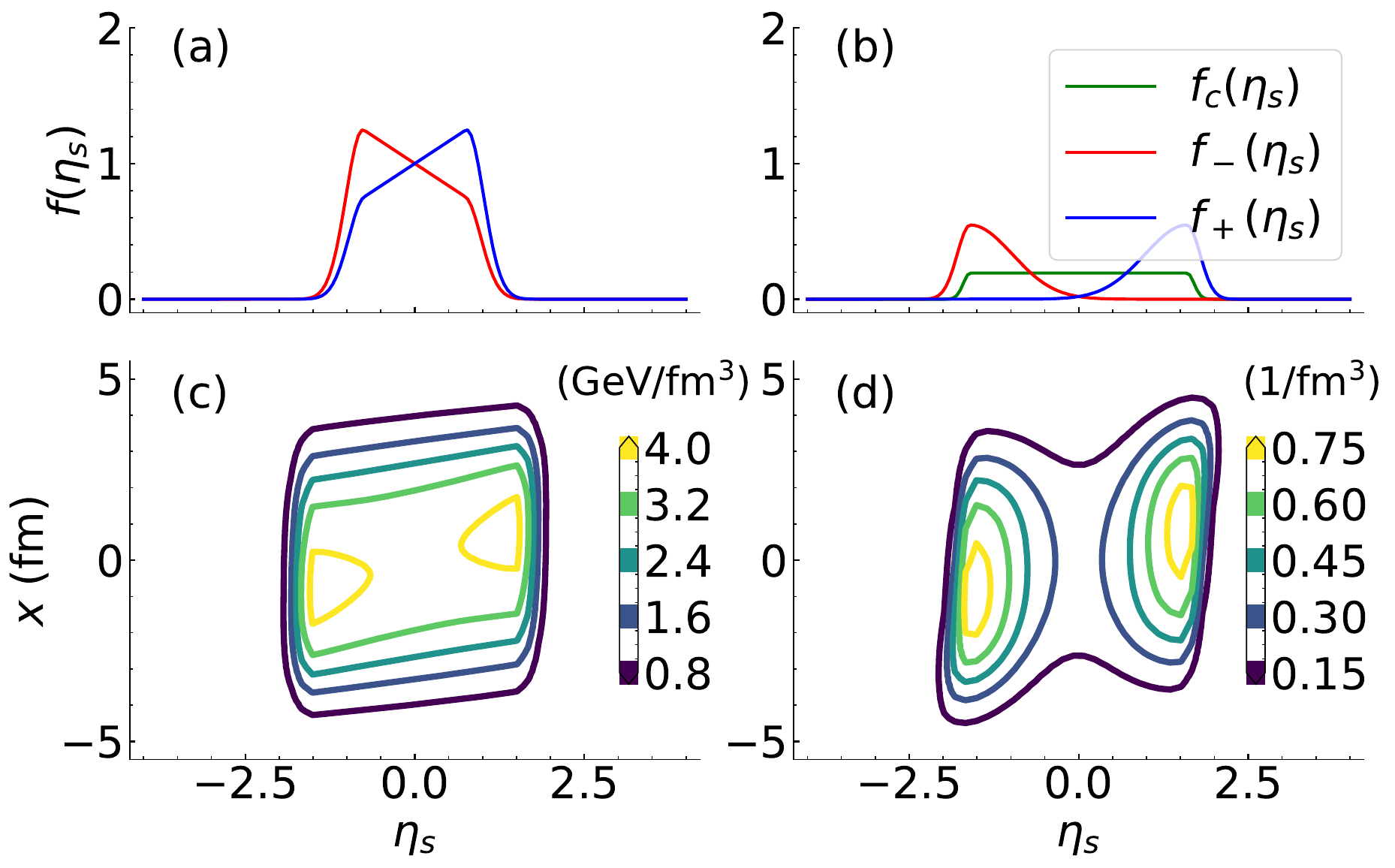}
\caption{%
    Longitudinal profiles of (a) entropy density and (b) baryon density, associated with the right-moving projectile (blue $f_+(\eta_s)$) and left-moving target (red $f_-(\eta_s)$), respectively, and the plateau component for baryon (green $f_c(\eta_s)$); the initial distributions of (c) energy density and (d) baryon density in the $x$-$\eta_s$ plane for 10-40\% Au+Au collisions at 19.6 GeV.}
    \label{fig:initial_condition}
\end{center}
\end{figure}

{\it \textbf{Model and setup.}---} The hydrodynamic stage starts at a constant $\tau_0$ with initial conditions constructed by extending the nucleus thickness function with parametrized longitudinal profiles \cite{suppl,Denicol:2018wdp}. Within a selected centrality class, the thickness functions are generated for $10^4$ fluctuating Monte-Carlo Glauber events, aligned up by their second-order participant plane angles, and then averaged to get smooth thickness functions of the projectile and target nuclei, $T_+(\xperp)$ and $T_-(\xperp)$, where $\xperp\equiv(x,y)$ denotes transverse coordinates. The same longitudinal profiles of entropy density $f^s_+(\eta_s)$ and $f^s_-(\eta_s)$ from Ref.~\cite{Denicol:2018wdp} are used to construct the entropy density $s(\tau_0,\xperp,\eta_s)$ (see Fig.~\ref{fig:initial_condition}(a)). We note that a ``tilt'' can be introduced to this type of 3-dimensional initial condition to explain the $v_1(y)$ for charged particles at 200 GeV \cite{Bozek:2010bi}.
These profiles were proposed because some studies suggested that emission at rapidities close to the beam rapidity is preferred \cite{Bialas:2004su,Bzdak:2009xq,Gazdzicki:2005rr,Bzdak:2009dr}. Such an initial entropy (energy) distribution drives the transverse expansion with a slight sideward component resulting in the $v_1(y)|_{y=0}$ with a negative slope of mesons (see Fig.~\ref{fig:initial_condition}(c)).

The initial baryon density, on the other hand, is highly related to the initial baryon stopping. The final net proton rapidity density is observed to have two peaks which get closer and eventually merge as the collision energy changes from 200 to 5 GeV \cite{Videbaek:2009zy} (see also Fig.~\ref{fig:yields}(b)). The two peaks move closer as the beam rapidity $y_\mathrm{b}$ decreases although the baryon rapidity loss also decreases \cite{BRAHMS:2009wlg}. The two peaks in $dN^{p-\bar{p}}/dy$ are generally attributed to the incoming nucleons losing rapidity after multiple collisions during the interpenetration stage (i.e., initial baryon stopping). In addition, baryon diffusion (i.e., hydrodynamic transport) also can evolve baryons from the high $\mu_B$ regions to the low $\mu_B$ regions \cite{Denicol:2018wdp,Du:2021zqz}. Thus, the final baryon stopping estimated from the net proton distribution has contributions from both the initial baryon stopping and from hydrodynamic transport; reducing the former can be compensated by increasing the latter and vice versa, which makes disentangling their effects challenging~\cite{Denicol:2018wdp,Du:2021zqz}.

To characterize the initial baryon stopping, we generalize the longitudinal baryon profiles from Ref.~\cite{Denicol:2018wdp} (see $f^B_+(\eta_s)$ and $f^B_-(\eta_s)$ in Fig.~\ref{fig:initial_condition}(b))  by introducing transverse dependencies according to the local asymmetry in the thickness functions of the projectile $T_+(\xperp)$ and target $T_-(\xperp)$ \cite{Shen:2020jwv}. We introduce the $\xperp$-$\eta_s$ correlation for baryon stopping based on the asymmetry between $T_+$ and $T_-$ with the parameter $r(\xperp)\equiv[(T_+-T_-)/(T_++T_-)]$ which is correlated to the net longitudinal momentum in the local center-of-mass frame \cite{Shen:2020jwv}.
We consider this collision kinematics by introducing an $\xperp$-dependent longitudinal shift in the baryon profiles, $f^B_-[\eta_L(\xperp)]$ and $f^B_+[\eta_L(\xperp)]$, with $\eta_L(\xperp){\,\equiv\,}[\eta_s-y_L^By_\mathrm{cm}(\xperp)]$; here $y_L^B$ is a parameter controlling how strong the longitudinal shift is, as a fraction of the center-of-mass rapidity $ y_\mathrm{cm}(\xperp)=\arctan\left[r(\xperp)\tanh(y_\mathrm{b})\right]$, and $y_b{\,=\,}{\rm arccosh}[\sqrt{s_{NN}}/(2 m_N)]$ is the incoming nucleon's beam rapidity \cite{Shen:2020jwv}. The shifts introduced by $\eta_L(\xperp)$ generate tilted peaks in baryon density in $x$-$\eta_s$ plane as shown in Fig.~\ref{fig:initial_condition}(d), describing the varying baryon stopping in the transverse plane. We shall show how such a generalized baryon profile helps to explain characteristic features of $v_1(y)$.

Simply using such initial baryon density profiles to fit the $dN^{p-\bar{p}}/dy$ would need a baryon peak at $\eta_s{\,>\,}0$ shifted towards $+x$ and at $\eta_s{\,<\,}0$ towards $-x$ \cite{Denicol:2018wdp}. The asymmetric distribution of baryon density with respect to the beam axis along $x$-direction, after the transverse expansion builds up \cite{Voloshin:1996nv}, would introduce a $v_1(y)$ with a positive slope for baryons strongly overshooting the data at all beam energies; e.g., the calculated $v_1(y)$ could be an order of magnitude stronger than the measurements using similar profiles without the longitudinal shift \cite{Shen:2020jwv}. Reducing the initial baryon density around midrapidity can suppress the baryon asymmetry and the consequently induced $v_1(y)$. However, choosing a large baryon conductivity $\kappa$ in the hydrodynamic phase cannot transport enough baryon charges to the midrapidity and obtain enough net protons, because the baryon diffusion current decays quickly \cite{Li:2018fow, Du:2021zqz}. Thus explaining the measured $v_1(y)$ of baryons  while also reproducing the $dN^{p-\bar{p}}/dy$ which has two peaks is extremely challenging \cite{Shen:2020jwv, Nara:2021fuu}; most of the previous studies have only dealt  with one of those aspects.

To address this issue, we introduce an $\eta_s$-independent plateau component, $f^B_c(\eta_s)$, in the baryon profiles, which has no preference for the projectile or for the target and thus contributes to the baryon density as $\propto (T_++T_-)f^B_c(\eta_s)$. We parametrize $f^B_c(\eta_s)$ using a plateau whose width is set equal to the distance between the peaks of $f^B_+(\eta_s)$ and $f^B_-(\eta_s)$, with two half-Gaussian tails (see Fig.~\ref{fig:initial_condition}(b)). The initial condition of baryon density is given by
\begin{align}\label{eq:b_prof}
    n(\tau_0,\xperp,\eta_s)&=\frac{N_B}{\tau_0}\Big\{f^B_-[\eta_L(\xperp)]T_-+f^B_+[\eta_L(\xperp)]T_+ +\nonumber \\
    & \cosh^{-2}[r(\xperp)]N_cf^B_c(\eta_s)\big(T_-+T_+\big)\Big\}\,,
\end{align}
where the $\cosh^{-2}[r(\xperp)]$ factor is used to suppress the plateau's contribution when $T_+$ and $T_-$ are less symmetric, and $N_c$ controls the contribution of the symmetric component $f^B_c$ to the net baryon charge relative to the $f^B_+(\eta_s)$ and $f^B_-(\eta_s)$ components. The overall normalization factor $N_B$ is fixed by fitting the measured $dN^{p-\bar{p}}/dy$.
We note that such a plateau previously was used to parameterize the energy profile \cite{Bozek:2010bi,Shen:2020jwv}, and was attributed to the binary collisions assumed to contribute in a symmetric way \cite{Bozek:2010bi}. Here we introduce such a plateau component for baryon density for the first time in a phenomenological study. Because it symmetrically contributes to the baryon density with respect to the beam axis, the $v_1(y)$ of baryons can be strongly suppressed overall, while enough net proton yields can still be achieved around midrapidity since the $f^B_c(\eta_s)$ component can contribute a flat net proton yields in rapidity.

Using different initial profiles for the baryon and energy densities implies that energy deposition and baryon stopping have different mechanisms, as shown in microscopic dynamical initialization models \cite{Shen:2017bsr,Shen:2022oyg,Du:2018mpf}. The plateau component in the baryon profile implies a new baryon stopping mechanism, which can perhaps be attributed to the string junction conjecture {as discussed for instance by Kharzeev \cite{Kharzeev:1996sq} and Sjostrand-Skands \cite{Sjostrand:2002ip}. In this picture, the baryon number within a nucleon would be associated with the string junction of a non-perturbative Y-shaped configuration of gluon fields. Kharzeev argued that there can be both single- and double-junction stopping in a nucleon-nucleon collision, whose cross sections are $\sigma^J{\,\sim\,}\mathrm{e}^{\alpha_1y_\mathrm{b}}[\mathrm{e}^{-\alpha_2(y+y_\mathrm{b})}+\mathrm{e}^{-\alpha_2(y-y_\mathrm{b})}]$ and $\sigma^{JJ}{\,\sim\,}\mathrm{e}^{-y_\mathrm{b}}$, respectively, where $\alpha_1{\,\approx\,}0.16, \alpha_2{\,\approx\,}0.58$ according to Regge theory and global data fit \cite{Kharzeev:1996sq,Brandenburg:2022hrp}. We note that the two rapidity-dependent terms in $\sigma^J$ correspond to the baryon stopping associated with the projectile and target, respectively, and can be connected to the $f^B_+$ and $f^B_-$ components in Eq.~\eqref{eq:b_prof}. On the other hand, the double-junction stopping $\sigma^{JJ}$ has no rapidity dependence and thus contributes a central plateau, which is exactly the property of the $f^B_c(\eta_s)$ component in Eq.~\eqref{eq:b_prof}. Furthermore, as $\sigma^{JJ}{\,\sim\,}\exp(-y_\mathrm{b}){\,\sim\,}(\sqrt{s_{NN}})^{-1}$, the plateau contribution increases when the beam energy decreases, which is also consistent with the trend in parameters used herein. There is less string junction stopping in peripheral collisions \cite{Kharzeev:1996sq}, a feature also captured by the $\cosh^{-2}[r(\xperp)]$ factor in  Eq.~\eqref{eq:b_prof}, which decreases when $|r|$ increases from central to peripheral collisions. The implementation of sophisticated string dynamics \cite{Sjostrand:2002ip} within the Lund string model \cite{Eden:1996xi,Andersson:1997xwk,Sjostrand:2003wg,Christiansen:2015yqa} has yielded similar baryon stopping profiles \cite{Fischer:2016zzs,Ferreres-Sole:2018vgo,Mohs:2019iee}.

{\it \textbf{Data selection.}---}
Before exploring the $v_1(y)$, we use the pseudorapidity distribution of charged particles $dN^\mathrm{ch}/d\eta$ and rapidity distribution of net protons $dN^{p-\bar{p}}/dy$ to constrain the longitudinal bulk dynamics of the simulations. The former imposes strong constraints on energy (and entropy) density, and  the latter on baryon density.

Because of the limited available measurements in large (pseudo-)rapidity windows, we focus on four beam energies in this work, $\sqrt{s_{NN}}{\,=\,}7.7,\,19.6,\,62.4,\,200\,$GeV (see Fig.~\ref{fig:yields}). The $dN^\mathrm{ch}/d\eta$ is measured by PHOBOS for 0-6\% Au+Au collisions at $\sqrt{s_{NN}}{\,=\,}19.6$, 62.4 and 200 GeV \cite{Back:2002wb}, and the $dN^{p-\bar{p}}/dy$ by BRAHMS for 0-5\% Au+Au at 200 GeV \cite{BRAHMS:2003wwg} and 0-10\% Au+Au at 62.4 GeV \cite{BRAHMS:2009wlg}. Based on the limiting fragmentation \cite{Benecke:1969sh} phenomenon observed in different collision systems at various beam energies for both charged particles \cite{Back:2002wb} and net protons \cite{BRAHMS:2009wlg}, we rescale the measurements of $dN^\mathrm{ch}/d\eta$ and $dN^{p-\bar{p}}/dy$ measured for 0-5\% Pb-Pb collisions at 8.77 \cite{NA50:2002edr,NA49:2010lhg} and 17.3 GeV \cite{NA49:1998gaz} by NA49 and NA50  to constrain our calculations for 0-5\% Au+Au at 7.7 and 19.6 GeV, respectively. At both beam energies, the $dN^{p-\bar{p}}/dy$ is rescaled to match the midrapidity $dN^{p-\bar{p}}/dy$ of 0-5\% Au+Au measured by STAR \cite{STAR:2017sal}, and the $dN^\mathrm{ch}/d\eta$ at 8.77 GeV is rescaled to constrain the model such that it can reproduce the midrapidity pion yields at 7.7 GeV \cite{STAR:2017sal}.

The $v_1(y)$ of identified particles is measured by STAR for 10-40\% Au+Au collisions \cite{STAR:2014clz, STAR:2017okv}, among which we focus on four species with the following kinematic cuts: pions $\pi^+$ and kaons $K^+$ within $p_T{\,>\,}0.2\, \gev$ and $p{\,<\,}1.6\, \gev$, protons $p$ within $0.4{\,<\,}p_T{\,<\,}2\,\gev$, and lambdas $\Lambda$ within $0.2{\,<\,}p_T{\,<\,}5\,\gev$ (see Fig.~\ref{fig:bes_v1}). The $v_1(y)$ of $\pi^+$ and $p$ is also measured in a larger rapidity window for midcentral Pb-Pb collisions at 17.3 GeV by NA49 \cite{NA49:2003njx}, which matches well with the data from STAR, and we use them for reference for Au+Au at 19.6 GeV.

\begin{figure}[!tb]
\begin{center}
\hspace{-.45cm}
\includegraphics[width=1.0\linewidth]{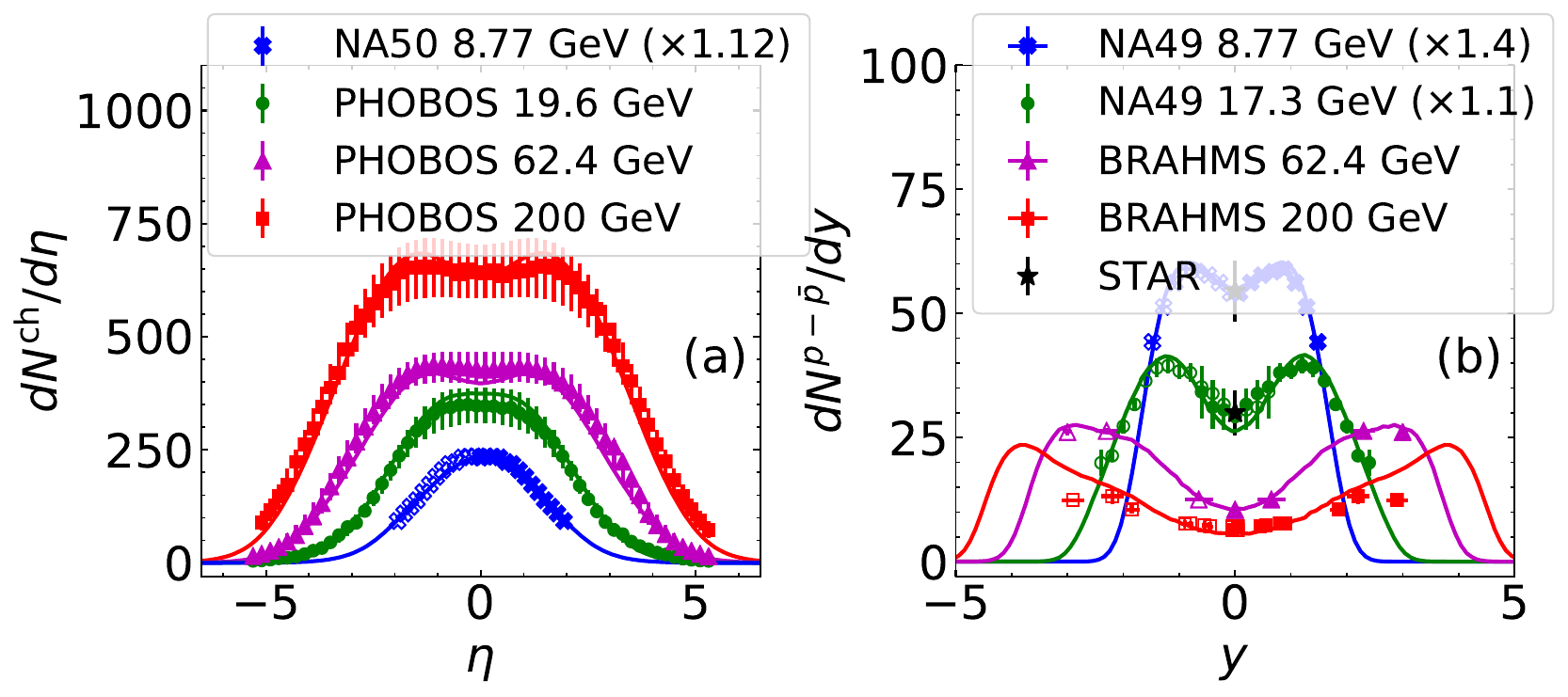}
\caption{%
    Comparisons for (a) $dN^\mathrm{ch}/d\eta$ and (b) $dN^{p-\bar{p}}/dy$ between the measurements (markers) and the hybrid model results (curves) for central Au+Au collisions at $\snn=7.7,\, 19.6,\, 62.4$ and 200 GeV.}
    \label{fig:yields}
\end{center}
\end{figure}


{\it \textbf{Results and discussion.}---}
The measurements of $dN^\mathrm{ch}/d\eta$ and $dN^{p-\bar{p}}/dy$ described above are used to calibrate the  theoretical calculations using NEOS-B (at vanishing strangeness and electric charge chemical potentials, $\mu_S {\,=\,} \mu_Q {\,=\,} 0$) for 0-5\% Au+Au collisions at 7.7, 19.6 and 200 GeV, and 0-10\% at 62.4 GeV (see Fig.~\ref{fig:yields}). Weak decay feed-down contribution to net protons is included in the calculation, as no weak decay correction is applied in the measurements at RHIC.
Admittedly, there still exists some mismatch between calculations and measurements, regarding the centralities and even collision systems in Fig.~\ref{fig:yields}, partly because of the lack of systematic measurements within the same centrality across beam energies. Nevertheless, considering the large error bars in the measurements, we think the longitudinal bulk dynamics is reasonably constrained by making use of the available data. The upcoming systematic BES-II measurements with large statistics will help to tighten constraints. 

The two observables in Fig.~\ref{fig:yields} are sensitive to the initial rapidity distributions but not to the tilted structure manifested in the $x$-$\eta_s$ plane. Our framework is then calibrated by the $v_1(y)$ of four identified hadron species in 10-40\% Au+Au collisions, as shown in Fig.~\ref{fig:bes_v1}, and the initial profiles in the $x$-$\eta_s$ plane are further constrained.
The $v_1(y)$ of mesons is mainly driven by the sideward pressure gradients, which is stronger toward $-x$ at $\eta_s{\,>\,}0$ and vice versa, originating from the tilted structure shown in Fig.~\ref{fig:initial_condition}(c). This structure successfully explained $v_1(y)$ of charged particles at the top RHIC energy \cite{Bozek:2010bi,Bozek:2011ua}. 
Here we note that the $v_1(y)$ of $\pi^+$ and $K^+$ are very similar in magnitude from our calculation using NEOS-B (left two columns of Fig.~\ref{fig:bes_v1}), and the similarity is consistent with the STAR measurements. Because of $\mu_S {\,=\,} \mu_Q {\,=\,} 0$ in NEOS-B, the difference between the $v_1(y)$ of $\pi^+$ and that of $K^+$ mainly originates from their mass difference in response to the underlying hydrodynamic flow velocity. 

We now focus on the $v_1(y)$ of baryons at various energies which has not been successfully explained within a hydrodynamic framework \cite{Singha:2016mna}. As shown in the right two columns of Fig.~\ref{fig:bes_v1}, our model successfully reproduces the $v_1(y)$ of $p$ and $\Lambda$ from the top RHIC energy to 7.7 GeV, and, in particular, the $dN^{p-\bar{p}}/dy$ at the same time. The rapidity-independent plateau component $f^B_c(\eta_s)$ in the initial baryon profile plays a critical role in achieving this agreement. As explained above, the plateau can reduce the asymmetry in baryon distribution with respect to the beam axis and thus the $v_1(y)$ of baryons, while producing enough $dN^{p-\bar{p}}/dy$ around midrapidity.
At high collision energies, such as 62.4 and 200 GeV, the plateau component of the net baryon density dominates the mid-rapidity region, resulting a flat and almost zero $v_1(y)$ of baryons within $|y|\lesssim1$.
Nevertheless, the components $f^B_-(\eta_s)$ and $f^B_+(\eta_s)$ at forward- and backward-rapidities on top of transverse expansion could still cause $v_1(y)$ with positive slope at $|y|\gtrsim1$, which is suppressed by the tilted peaks in our calculations. Again, the tilted peaks are introduced to account for the varying baryon stopping in the transverse plane, thus measuring $v_1(y)$ in a large rapidity window would help to constrain it.

\begin{figure}[!tb]
\begin{center}
\hspace{-.5cm}
\includegraphics[width=0.98\linewidth]{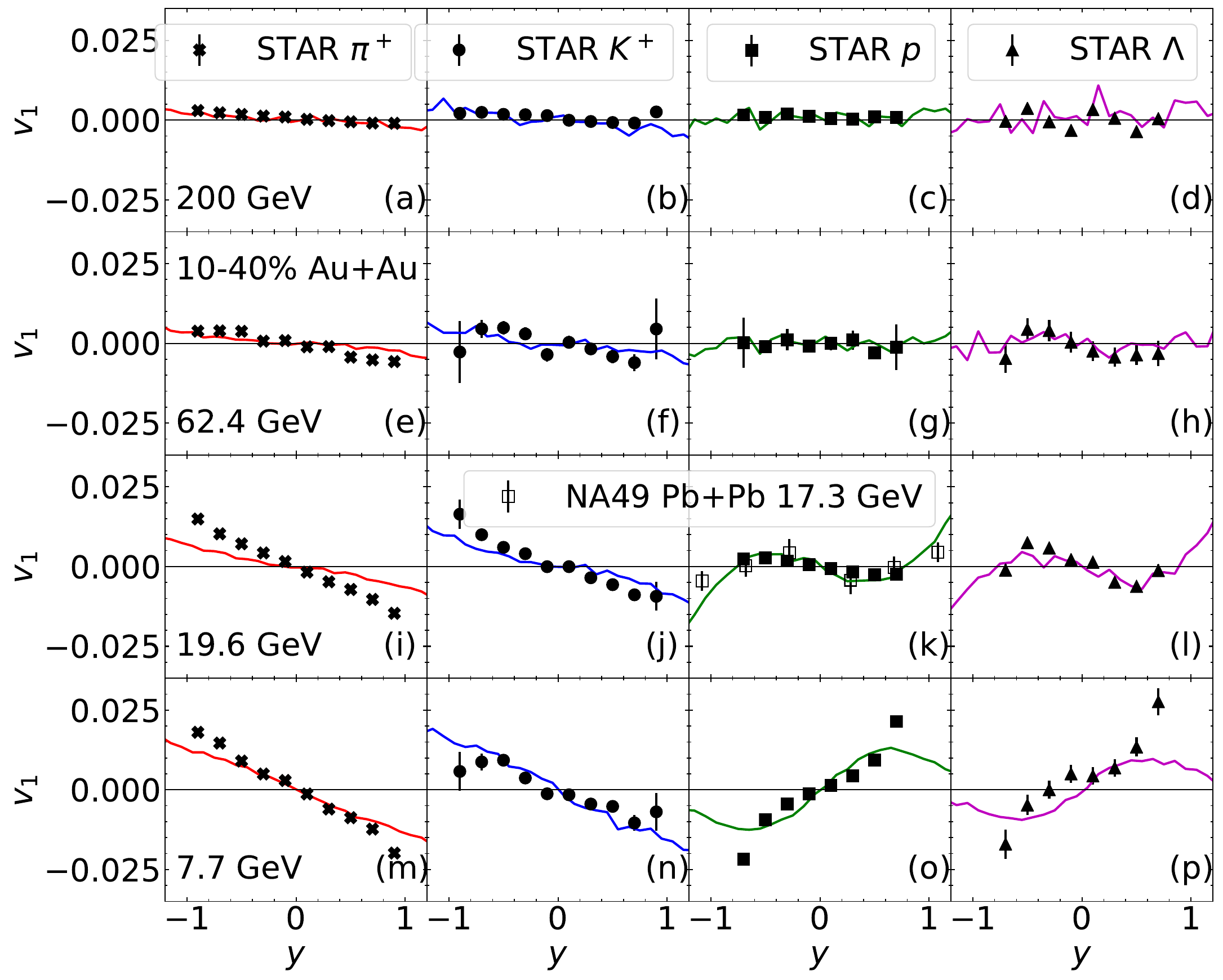}
\caption{%
    Comparison between measurements (markers) and calculations (lines) for the $v_1(y)$ of $\pi^+$, $K^+$, $p$ and $\Lambda$ (from left to right) in 10-40\% Au+Au collisions at 7.7, 19.6, 62.4 and 200 GeV (from bottom to top).}
    \label{fig:bes_v1}
\end{center}
\end{figure}

At 19.6 GeV, the $v_1(y)$ of baryons and especially its feature of a cubic rapidity dependence (``wiggle'') is nicely reproduced in our model when comparing to the STAR and NA49 measurements  (see Fig.~\ref{fig:bes_v1}(k,l)). Such a wiggle is more pronounced for $\Lambda$'s $v_1(y)$ as shown in both our calculations and the STAR measurements. The baryon distribution illustrated in Fig.~\ref{fig:initial_condition}(d) with tilted peaks naturally generates the wiggle of $v_1(y)$ for baryons: Focusing on $y{\,\gtrsim\,}0$ rapidity regions, more baryon is distributed at $x{\,<\,}0$ near midrapidity while at $x{\,>\,}0$ at forward rapidities, and the former generates negative $v_1$ while the latter positive $v_1$ after the transverse expansion develops. We emphasize that to achieve this, the two peaks in Fig.~\ref{fig:initial_condition}(d) need to be well separated which again necessitates the presence of the $f^B_c(\eta_s)$ component: Without $f^B_c(\eta_s)$, to generate enough net protons around midrapidity, $f^B_-(\eta_s)$ and $f^B_+(\eta_s)$ would need larger inward tails; consequently the two peaks in Fig.~\ref{fig:initial_condition}(d) would merge, and thus the right tilted peak would not distribute more baryon at $x{\,<\,}0$ near midrapidity, which results in $v_1(y)$ with a positive slope with no cubic rapidity dependence. On the other hand, without $f^B_c(\eta_s)$, $f^B_-(\eta_s)$ and $f^B_+(\eta_s)$ would need higher peaks to reproduce the measured $dN^{p-\bar{p}}/dy$, resulting in much stronger $v_1(y)$ that overshoots the data. As shown in Figs.~\ref{fig:v1_comp}(a-d), the framework is retuned to fit the $dN^{p-\bar{p}}/dy$ without $f^B_c(\eta_s)$, and then the $v_1(y)$ of baryons becomes much stronger with a maximum value $|v_1|_\mathrm{max}\approx 0.07$ around $|y|=1$ and the wiggle simply disappears. We note that $|v_1|_\mathrm{max}$ here is about half of that in Ref.~\cite{Shen:2020jwv}, because the baryon diffusion is included here which can smooth baryon asymmetry and thus reduce $v_1$ but not strongly enough. 

At 7.7 GeV, the consequences are illustrated when the baryon peaks merge because of the smaller beam rapidity: Such a baryon distribution with transverse expansion naturally gives $v_1(y)$ with a positive slope (Fig.~\ref{fig:bes_v1}(o,p)); however, in this case, although the tilted peaks do not generate the cubic rapidity dependence in $v_1(y)$, it still helps to reduce the positive slope, which makes $v_1(y)$ around midrapidity more consistent with the STAR measurements. Again, we emphasize the essence of the $f^B_c(\eta_s)$ component in suppressing $v_1(y)$ and generating the relatively small $v_1(y)$ at 7.7 GeV. It is noticeable that the $v_1(y)$ of baryons we obtain starts decreasing at $|y|\gtrsim0.7$, while the measured value keeps increasing. This difference may originate from the interactions between the spectators and the fireball which are expected to be important at low beam energies \cite{Zhang:2018wlk,Nara:2021fuu} but not considered in this work. 

\begin{figure}[!tb]
\begin{center}
\hspace{-.5cm}
\includegraphics[width=0.98\linewidth]{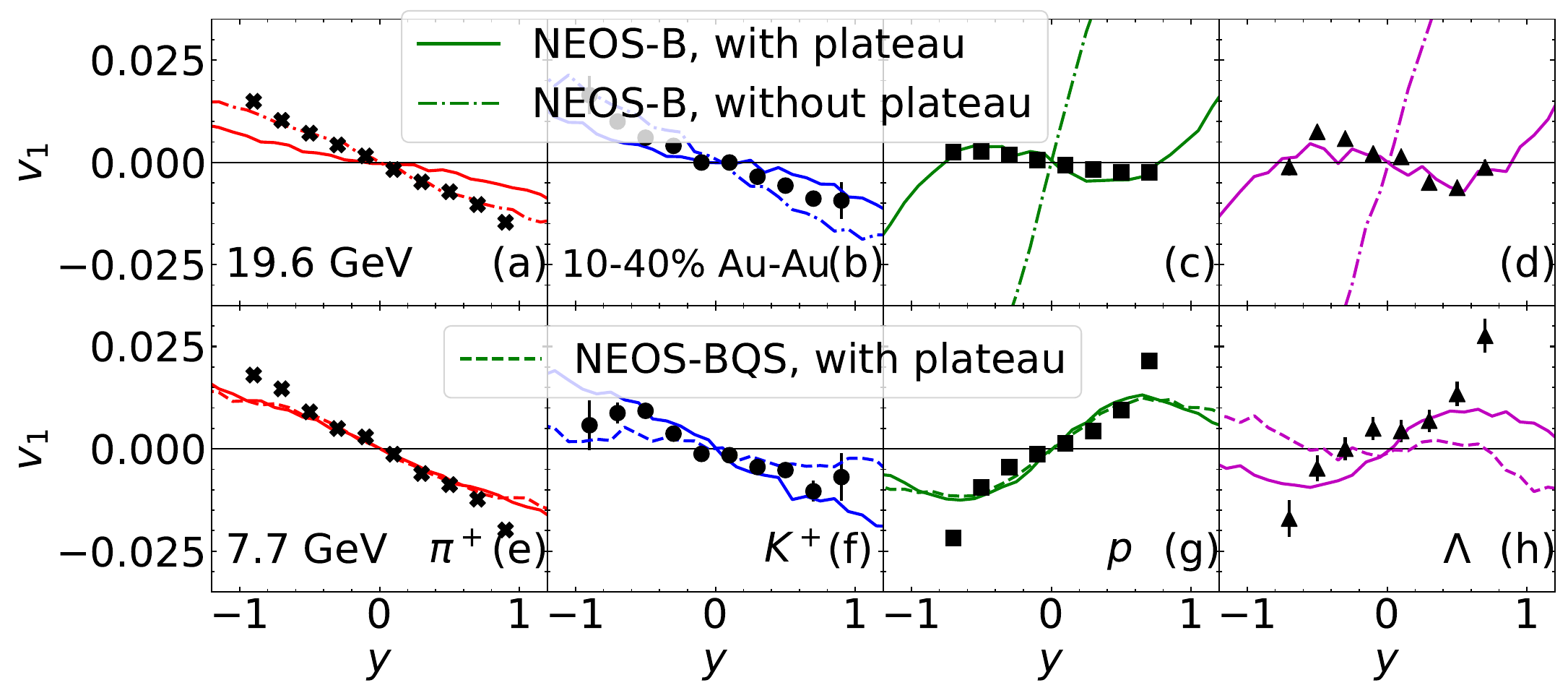}
\caption{%
    Comparison of $v_1(y)$ in 10-40\% Au+Au collisions from Fig.~\ref{fig:bes_v1} using NEOS-B with the plateau (solid) to the results at 19.6 GeV using NEOS-B but without plateau (dot-dashed), and to those at 7.7 GeV with plateau using NEOS-BQS (dashed).}
    \label{fig:v1_comp}
\end{center}
\end{figure}

Finally, we explore how the $v_1(y)$ of identified particles can provide information of the chemistry of the nuclear matter, by comparing results using NEOS-B to those using NEOS-BQS (assuming strangeness neutrality $n_S {\,=\,} 0$ and fixed electric charge-to-baryon ratio $n_Q{\,=\,}0.4 n_B$) at 7.7 GeV in Figs.~\ref{fig:v1_comp}(e)-(h). The latter imposes local strangeness neutrality and thus strongly couples the production of $K^+$ and $\Lambda$ which carry the chemical potentials $\mu_Q+\mu_S$ and $\mu_B-\mu_S$, respectively. For example, at $y{\,>\,}0$, because the sideward pressure gradient drives more $K^+$ towards $-x$-direction, more $\Lambda$s come alongside to assure strangeness neutrality, acting against the production of $\Lambda$ towards $+x$-direction due to the asymmetric baryon distribution. This suppresses the $v_1$ of $\Lambda$s around midrapidity, and even alters its sign beyond $|y|\gtrsim0.6$ (Fig.~\ref{fig:v1_comp}(h)). Similarly, the $v_1$ of $K^+$ in turn gets suppressed because of $\Lambda$s. As a results, requiring strangeness neutrality destroys the similarity indicated by the measurements in $v_1$ of $\pi^+$ and $K^+$, and that in $p$ and $\Lambda$ at 7.7 GeV. We have also checked that using NEOS-BQS only slightly changes $v_1(y)$ at 19.6 GeV and higher beam energies. This suggests that to properly describe the dynamical evolution around 10 GeV and lower center-of-mass energies, the simulations propagating multiple charges with EoS at finite $\mu_{B, Q, S}$ without constraints imposed by NEOS-B and NEOS-BQS would be necessary. This low beam energy region is covered by the BES-II program, and thus the upcoming rapidity-dependent $v_1(y)$ measurements can probe the initial distributions and EoS of multiple charges.

{\it \textbf{Summary.}---}
Using a (3+1)-dimensional hybrid framework with parametric initial conditions, we successfully reproduce the measured rapidity and beam energy dependence of the directed flow $v_1(y)$ of identified particles from 7.7 to 200 GeV.
Since a smooth cross-over EoS is used, the observed sign change in the net protons' $dv_1/dy|_{y=0}$ around 10-20 GeV collision energy is not an evidence of the first-order phase transition of the QCD matter, but rather a consequence of the initial baryon stopping.

In the initial baryon profile, we introduce a rapidity-independent plateau component  for the first time in a phenomenological study. It is essential in reducing the asymmetric baryon distribution with respect to the beam axis and thus preventing strong $v_1(y)$ of baryons, and meanwhile in generating enough net protons around midrapidity to be consistent with the measurements. Especially, together with the tilted baryon peaks in the reaction plane which account for the transversely varying baryon stopping, it explains the $v_1(y)$ of baryons with double sign change in $dv_1(y)/dy$ at 19.6 GeV and naturally generates that with a positive $dv_1(y)/dy$ around midrapidity at 7.7 GeV. Thus our work demonstrates that the $v_1(y)$ of identified hadrons and $dN^{p-\bar{p}}/dy$ together can strongly constrain the initial baryon stopping and especially its transverse plane dependence.

The plateau component in the initial baryon profile describes a special baryon stopping mechanism which may originate from the string junction stopping conjecture proposed more than two decades ago \cite{Kharzeev:1996sq,Sjostrand:2002ip}. Recently this proposal saw a resurgence of interest in the community \cite{Shen:2022oyg,Brandenburg:2022hrp} and ideas for new measurements testing it have been proposed \cite{Brandenburg:2022hrp}. We have shown that the plateau component is essential to explain these rapidity-dependent particle production and directed flow measurements.  Nevertheless, it is important to implement this baryon stopping component in more comprehensive framework and carry out event-by-event simulations~\cite{Shen:2022oyg}. Note that Bayesian model selection facilitated with the upcoming BES-II measurements can quantify the support for models implementing various initial baryon stopping mechanisms, including the consequences of string junction stopping scenarios. 

Finally, our study indicates that the rapidity-dependent directed flow measurements of identify strangeness particles around 10 GeV and lower center-of-mass energies can probe the EoS at finite chemical potentials. Hence the upcoming $v_1(y)$ measurements at BES-II, especially at its fixed-target program, could strongly improve our understanding of the nuclear matter in the vicinity of the hypothetical QCD critical point.

{\it \textbf{Acknowledgements.}---}
This work was supported in part (L.D., S.J., C.G.) by the Natural Sciences and Engineering Research Council of Canada, and in part (C.S.) by the U.S. Department of Energy, Office of Science, Office of Nuclear Physics, under DOE Award No. DE-SC0021969 and DE-SC0013460. C.S. also acknowledges support from a DOE Office of Science Early Career Award.

\bibliography{ref}
\onecolumngrid

\appendix*

\section*{Supplemental Material}
In this Supplemental Material we describe the framework and parameters used for this study. In the hydrodynamic stage, we propagate the energy-momentum tensor and net baryon current with the dissipative effects from the shear stress tensor and net baryon diffusion current \cite{Denicol:2018wdp}. These dissipative currents are described by the Israel-Stewart-like equations given by the second-order Denicol–Niemi–Molnar–Rischke theory \cite{Denicol:2012cn}. The specific shear viscosity $\eta/s$ has both temperature $T$ and baryon chemical potential $\mu_B$ dependence \cite{Shen:2020jwv}, which increases rapidly at low $T$ and high $\mu_B$ regions (see Fig.~4 of Ref.~\cite{Shen:2020jwv}):
\begin{equation}
    (\eta/s)(T, \mu_B) = (\eta/s)_0 f_T(T) \frac{e+P}{Ts} f_{\mu_B}(\mu_B)\,,
\end{equation}
where
\begin{equation}
    f_T(T) = 1 + 1.2 \left(\frac{\frac{T}{1\,\mathrm{GeV}} - 0.165}{0.065}\right) \theta \left( 0.165 - \frac{T}{1\,\mathrm{GeV}} \right)\,,
\end{equation}
and
\begin{equation}
    f_{\mu_B}(\mu_B) = 1 + 0.9 \left(\frac{\mu_B}{0.6\,\mbox{GeV}}\right)^{1.2}\,;
\end{equation}
for the overall coefficient, we use $(\eta/s)_0=0.1$. The baryon diffusion conductivity $\kappa$ is obtained from the Boltzmann equation in the relaxation time approximation in the massless limit \cite{Denicol:2018wdp}:
\begin{equation}
    \kappa = \frac{C_B}{T} n_B \left(\frac{1}{3} \coth\left(\frac{\mu_B}{T}\right) - \frac{n_B T}{e + P}\right)\,,
\end{equation}
with $C_B=0.1$. We use an EoS which smoothly interpolates between the result obtained using lattice QCD EoS at high $T$, and that associated with a hadron resonance gas one at low $T$ \cite{Monnai:2019hkn}. We explore two limits of the full four-dimensional EoS phase space in ($T$, $\mu_B$, $\mu_S$, $\mu_Q$): vanishing strangeness and electric charge chemical potentials, $\mu_S {\,=\,} \mu_Q {\,=\,} 0$ (NEOS-B) and nonzero potentials with strangeness neutrality $n_S {\,=\,} 0$ and fixed electric charge-to-baryon ratio $n_Q{\,=\,}0.4 n_B$ (NEOS-BQS) \cite{Monnai:2019hkn}. When the system expands and cools, we switch to a kinetic transport description and use UrQMD \cite{Bass:1998ca,Bleicher:1999xi}. The particlization process is carried out on a hyper-surface at a constant energy density $e_\mathrm{sw}{\,=\,}0.26$ GeV/fm$^3$ \cite{Huovinen:2012is, Shen:2020jwv}. The off-equilibrium effects are considered when the hadrons are sampled on the freezeout surface using the Cooper-Frye prescription \cite{Cooper:1974mv}. The framework used in this study is open source and accessible on GitHub \cite{iebe}.

At  Beam Energy Scan (BES) energies, the hydrodynamization of the collision system has a non-trivial space and time dependence. In principle, initializing the hydrodynamic evolution requires a microscopic dynamical description involving collisions between nucleons, between partons,  and  demands knowledge of the energy deposition and baryon-stopping mechanisms. Recently, some progress on such dynamical initialization has been made, and some success has been achieved in describing the dynamics along the beam direction at both RHIC and LHC energies using the same framework \cite{Shen:2017bsr,Shen:2022oyg}. In this work, instead of using such a dynamical initialization method, we construct parametric initial conditions and initialize the hydrodynamic description at a constant $\tau_0$ with the Bjorken flow. With the initial conditions described below, the hydrodynamic evolution starts at the initialization time $\tau_0=3.6, 1.8, 1.0, 1.0$ fm$/c$ from 7.7 to 200 GeV \cite{Shen:2020jwv}. The pressure gradients drive the hydrodynamic expansion and collective flows. In this way, we can isolate the initial profiles favored by the data, and this will lead to insight into the dynamics of initial energy deposition and of baryon-stopping. 

The initial condition of entropy density is given by \cite{Denicol:2018wdp}
\begin{equation}
    s(\tau_0,\xperp,\eta_s)= s_0 [f^s_-(\eta_s)T_-+f^s_+(\eta_s)T_+]\,,
\end{equation}
where $s_0$ is an overall normalization factor tuned to fit the charged particle multiplicity. Here $f^s_\pm(\eta_s)$ are the longitudinal profiles associated with the projectile and target, given by
\begin{align}
&f^s_\pm(\eta_s) = \theta(\eta_\mathrm{max} - \vert \eta_s \vert) \left(1 \pm \frac{\eta_s}{\eta_\mathrm{max}} \right)  \\
& \,\times \left[\theta(\vert \eta_s \vert - \eta^s_0) \exp\left(-\frac{(|\eta_s| - \eta^s_0)^2}{2\sigma_{\eta,s}^2} \right) + \theta(\eta^s_0 - \vert \eta_s \vert) \right]\notag\,,
\end{align}
where the maximum extension in $\eta_s$ is chosen to be the beam rapidity, i.e., $\eta_\mathrm{max}=y_\mathrm{b}$ (see Fig.~\ref{fig:initial_condition}(a)). 

The initial profile of baryon density is given by Eq.~\eqref{eq:b_prof}, and $f^B_\pm(\eta_s)$ are chosen as
\begin{align}
f^{B}_\pm (\eta_s) &= N \left[\theta(\eta_s - \eta^{B, \pm}_0) \exp\left(-\frac{(\eta_s - \eta^{B, \pm}_0)^2}{2\sigma_{B,\mp}^2} \right) \right. \notag \\
 &\left. + \theta(\eta^{B, \pm}_0 - \eta_s) \exp\left(-\frac{(\eta_s - \eta^{B, \pm}_0)^2}{2\sigma_{B, \mp}^2} \right) \right]\,,
\end{align}
where the normalization factor $N$ ensures $\int f^B_\pm(\eta_s)d\eta_s=1$, and $\eta^{B, \pm}_0$ control locations of the peaks. The central plateau component is parameterized as
\begin{equation}
    f^B_c(\eta_s)=N'\exp\left[-\frac{(|\eta_s|-\eta_0^B)^2}{2{\sigma_{\eta,B}}^2}\theta(|\eta_s|-\eta_0^B)\right]\,,
\end{equation}
where $N'$ is a normalization factor ensuring $\int f^B_c(\eta_s)d\eta_s{\,=\,}1$. Here $\eta_0^B$ is the width of the plateau and $\sigma_{\eta,B}$ is the width of its two half-Gaussian tails. We set $\eta^{B, +}_0=-\eta^{B, -}_0=\eta_0^B/2$ in this work (see Fig.~\ref{fig:initial_condition}(b)). Parameters in the initial profiles are determined by the measured $dN^{p-\bar{p}}/dy$ and $dN^\mathrm{ch}/d\eta$, together with $v_1(y)$ of identified particles, and are listed in Table \ref{tab:init}. 

\begin{table}[!tb]
\centering
\begin{tabular}{ccccccccccccc}
\hline
\hline
$\sqrt{s_\mathrm{NN}}$ GeV & $s_0$ & $\eta^s_0$ & $\sigma_{\eta,s}$ & $N_B$ & $\sigma_{B,+}$ & $\sigma_{B,-}$ & $y_L^B$ & $N_c$ & $\eta^B_0$ & $\sigma_{\eta,B}$ \\
\hline
7.7 &  2.45 & 0.75 & 0.17 & 0.8  & 0.15 & 0.58 & 0.2  & 0.6  & 2.3 & 0.07 \\
19.6 &  5.85 & 1.5  & 0.25 & 0.6  & 0.19 & 0.62 & 0.3  & 0.65 & 3.2 & 0.1 \\
62.4 &  11.0 & 2.3  & 0.28 & 0.65 & 0.25 & 0.95 & 0.25 & 0.51 & 5.4 & 0.22 \\
200 &  16.0 & 2.5  & 0.58 & 0.55 & 0.3  & 1.15 & 0.1  & 0.41 & 7.0 & 0.25 \\
\hline
\hline
\end{tabular}
\caption{Parameters of initial profiles at four beam energies.}
\label{tab:init}
\end{table}

\end{document}